\documentclass[12pt]{article}
\usepackage{epsfig,epsf,rotating}
\begin{document}

\title{Self-Organized Criticality in Particle Production}

\author{A.Paramonov, A.Rostovtsev\\
\\
{\it Institute f. Theoretical and Experimental Physics, ITEP,} \\
{\it Moscow, Russia}
}
\date{ }
\maketitle
\begin{center}
Abstract
\end{center}
Self-Organized Criticality paradigm is a plausible picture for
hadron production. A power-law behavior of   
hadron transverse momentum spectra and an approximate scaling
observed for different hadrons in high energy hadronic collisions are
discussed. 
\vspace*{1.0cm}

Multi-particle production in high energy hadronic collisions is
understood in Quantum Chromodynamics as a result of irradiation and
recombination of many strongly interacting quarks and gluons. 
A system of hadronizing soft partons is an open dynamical system 
(the number of partons is not conserved) with many
degrees of freedom which is far from equilibrium.
Even though we can calculate exactly an elementary parton-parton
interaction, a complexity of the system of hadronizing partons 
doesn't allow the extension of these calculations to predict
particle states,
in which the vast majority of the particles
are produced with low transverse momenta~$(P_t)$.
In addition, cascading decays of produced hadronic resonances make 
the picture even more complex. In practice, particle production
in high energy hadronic collisions is modeled using
statistical techniques.

The seemingly structureless multi-particle final states,  
however, are found to display regularities, namely, long-range
particle-particle correlation and self-similarity in the particle spectra.
Interestingly, a similar behavior is found for a broad class of phenomena
including turbulence, earthquakes, forest fires, cloud formation,~{\it
etc.} The phenomena from this class are described by power-law
distributions.
To explain the occurrence of power-laws in nature
P.Bak~{\it et al}~\cite{Bak} have proposed a Self-Organized
Criticality~(SOC) paradigm.
In the SOC the dissipative systems spontaneously
evolve into barely stable critical states, with long range correlations. 
The critical state is distinguished from the fully deterministic states 
by the response of a system to an external perturbation.
A reaction of deterministic system is described by a characteristic
response at a given scale depending on the system's parameters.
The distribution of the resulting events is narrow and is well described
by an average value.
For a critical system, the same perturbation can lead to an event of any
scale. Although large power events are comparatively rare, 
the mechanism is the same to explain the rare large
events and the smaller, more common ones.
This scaling behavior is described by a power
law distribution.

In it's relatively short history SOC patterns have been found in myriads
of fields including biology, sociology, computer science, economics, {\it
etc}. The SOC models are widely used in modern physics to describe
processes, such as occurrence of solar flares, earthquakes, and
windstorms.
The SOC gives rise to fractal dimensions of natural
objects. The ubiquitous ``fingerprints" of the SOC states are
a power-law distribution of the events and 
long-range correlations.
In high energy physics the
long-range correlations in particle production have been found and
studied in details before~\cite{correlations}. 
In the present paper we discuss a power-law behavior of
the transverse momentum spectra of different hadrons measured in
collider experiments.

At low transverse momenta an exponential behavior of the hadronic spectra
has been observed. It was interpreted using a
thermodynamic analogy,
with single particles distributed within a hot hadronizing matter
according to R.~Hagedorn~\cite{Hagedorn}
\begin{equation}
E\left.\frac{d^3\sigma}{d^3p}\right|_{y=0} \sim 
\sqrt{m^2+P_t^2}\cdot exp(-\frac{\sqrt{m^2+P_t^2}}{T})\,,
\label{Boltzmann}
\end{equation}
where $m$ is a particle mass and $T$ is characteristic temperature of
the interaction\footnote{Since the most of the data on hadron
production are available for central rapidities~($y_{lab}\approx 0$)
we consider the invariant differential cross sections for central 
rapidity only.}. 
The data on high-$P_t$~particle production show a significant deviation from
the exponential form towards a power-law
\begin{equation}
      E\frac{d^3\sigma}{d^3p} \sim \left(\frac{1}{P_t}\right)^n\,,
\label{pQCD}
\end{equation}
anticipated by the perturbative QCD calculations for parton scattering.
These calculations, however, fail in the low-$P_t$ region.
An alternative statistical hadronization 
model~\cite{Becattini_model}
based on the formation and decay of pre-hadronic clusters was found to
give a good data description within a broad range 
of values of hadron~$P_t$, though
it fails to describe the interactions with collision energy larger 
than~$30~GeV$~\cite{Becattini_30}.  
To describe high energy data within the whole $P_t$ interval 
Ga\'{z}dzicki~{\it et~al}~\cite{Gazdzicki} have suggested
a new statistical
parameterization of hadron spectra which approximates the 
exponential form 
at low-$P_t$ and the power-law at high-$P_t$ 
\begin{equation}
 E\left.\frac{d^3\sigma}{d^3p}\right|_{y=0} \sim 
\left(\frac{1}{\sqrt{m^2+P_t^2}}\right)^{D}\,.
\label{Gazdzicki}
\end{equation}

In practice, to describe the hadron spectra an empirical 
power-law function is widely used
\begin{equation}
 E\left.\frac{d^3\sigma}{d^3p}\right|_{y=0} \sim 
\left(\frac{1}{P_0+P_t}\right)^{D}\,.
\label{Power-law}
\end{equation}
The form~(\ref{Power-law}) is a very close approximation of
the following expression
\begin{equation}
 E\left.\frac{d^3\sigma}{d^3p}\right|_{y=0}
 \sim \frac{\sqrt{m^2+P_t^2}}
{(T_0+(q-1)\sqrt{m^2+P_t^2}\,)^{\frac{q}{q-1}}}
\label{Non-extensive}
\end{equation}
obtained in a framework of
non-extensive statistical mechanics~\cite{Tsallis,Beck}.
In~(\ref{Non-extensive}) the $T_0$ is a temperature of hadronizing matter,
$m$ is a hadron mass and $q$ is an entropic index. For $q\rightarrow 1$
the expression~(\ref{Non-extensive}) reduces to the Hagedorn 
form~(\ref{Boltzmann}).
The non-extensive statistical mechanics generalizes the standard
thermodynamic approach for a presence of a long-range strong interaction
and is deeply related to the SOC~\cite{Tsallis_SOC}.
It was shown~\cite{Bediaga,Beck} that non-extensive statistical model
describes the hadron spectra in $e^+e^-$ interaction.
It is of interest to investigate which parameterization 
from~(\ref{Gazdzicki}) to~(\ref{Non-extensive}) is preferred by the
high energy hadron collider data.
\begin{figure}[h]
\begin{center}
\hspace*{-5.0cm}
\epsfig{
file=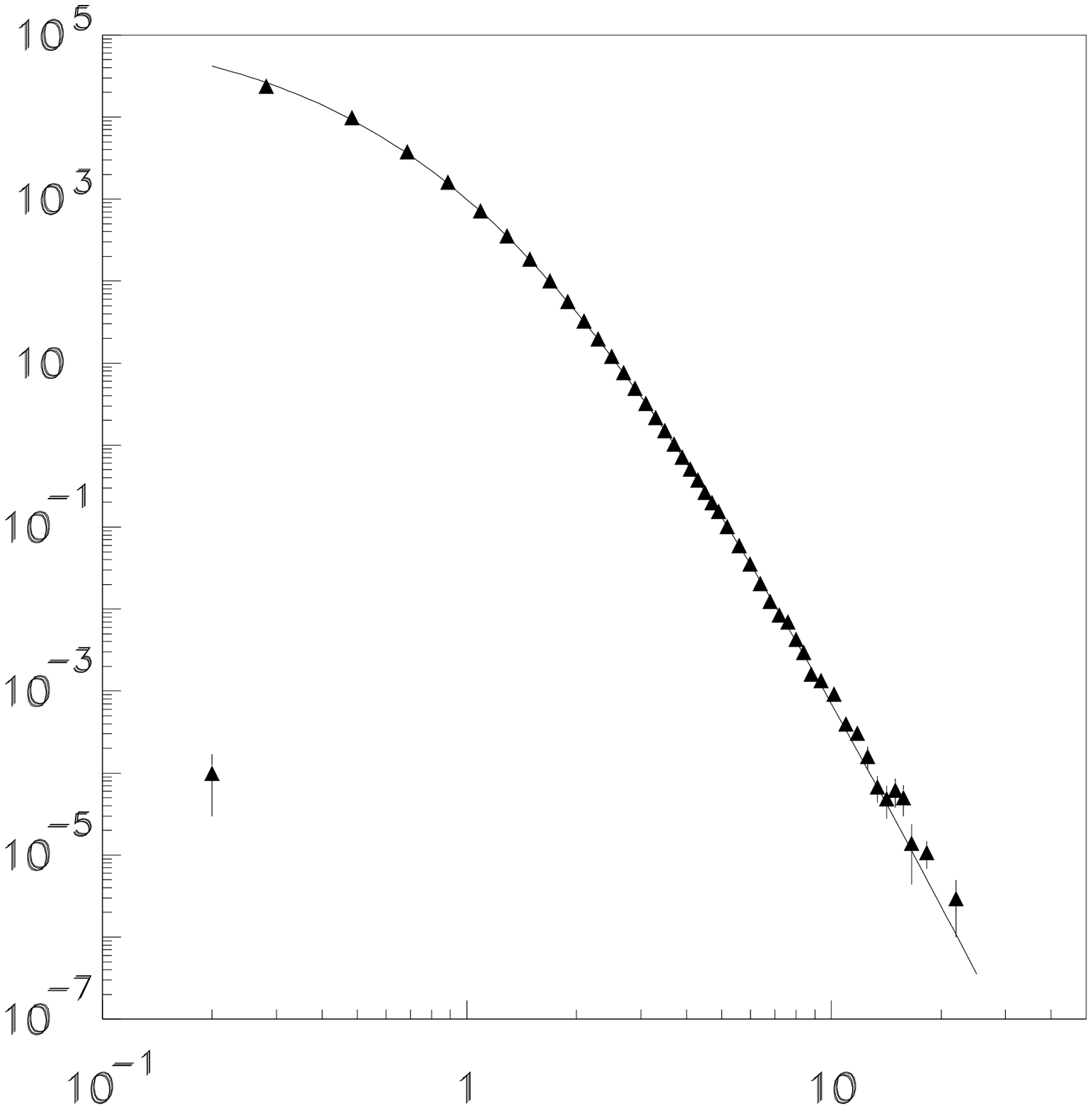,
        height=8.0cm,width=7.0cm,angle=0}
\put(-148,63){\large\bf $h^+ + h^-$}
\put(-40,185){\Large\bf a}
\put(-150,110){\Large UA1}
\put(-165,85){\small\bf $\sqrt{s}=560~GeV$}
\put(-80,-10){\large\bf $P_t~[GeV]$}
\put(-225,60){\begin{sideways}\Large\bf
$E\left.\frac{d^3\sigma}{d^3P}\right|_{y=0} [\mu{b}/GeV^2]$\end{sideways}}

\vspace*{-9.6cm}\hspace*{8.2cm}
\epsfig{
file=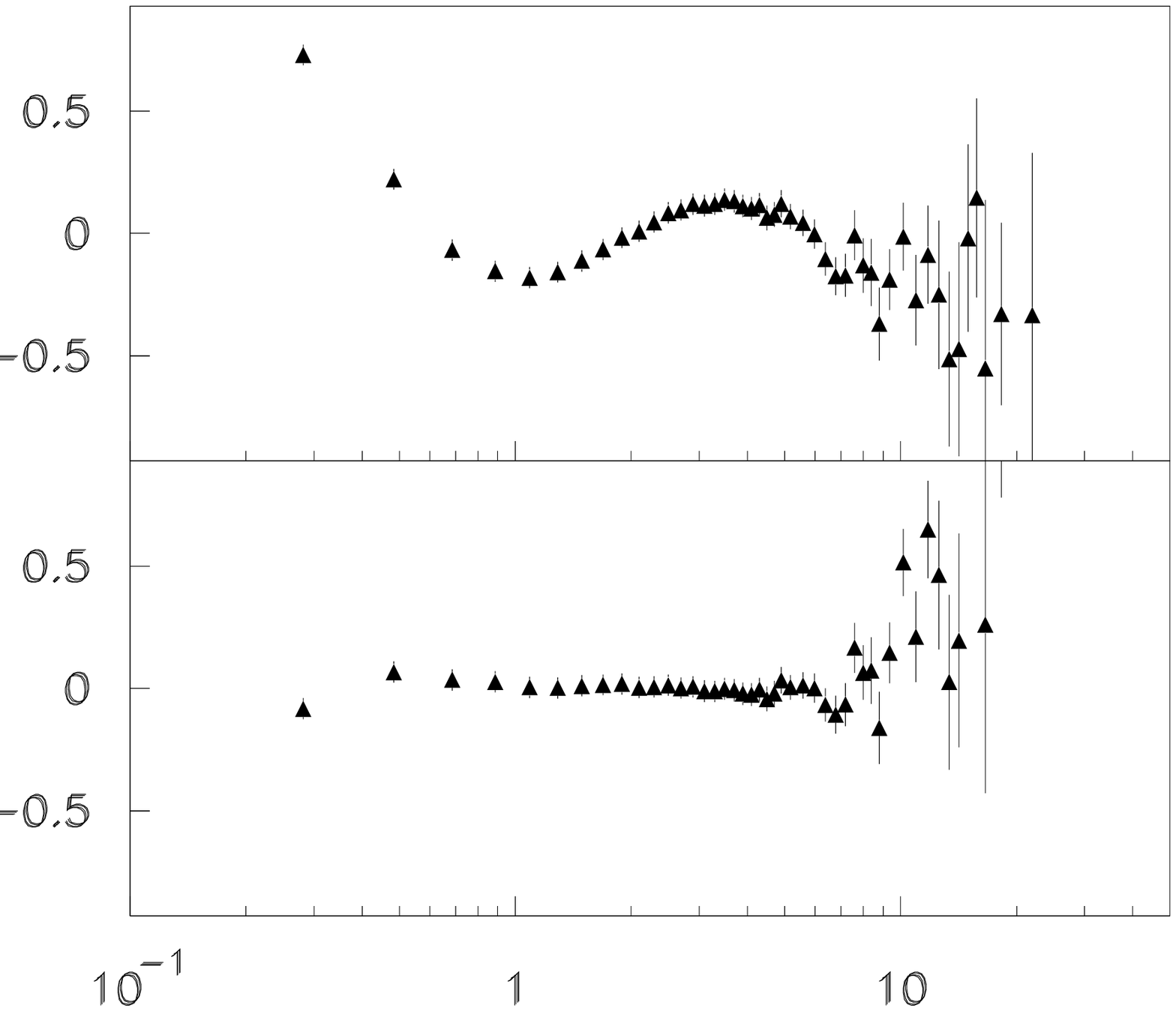,
        height=9.1cm,width=6.0cm,angle=0}
\put(-33,185){\Large\bf b}
\put(-147,127){\normalsize\bf $\sim(\sqrt{m^2+P_t^2})^{-D}$}
\put(-33,95){\Large\bf c}
\put(-147,37){\normalsize\bf $\sim(P_0+P_t)^{-D}$}
\put(-80,-10){\large\bf $P_t~[GeV]$}
\put(-190,65){\begin{sideways}\Large (data - fit)/fit \end{sideways}}
\caption{
The measured charged particle $P_t$ distribution with an overlayed
the power-law curve~(a), and a
relative difference between the data and the fit results using power-law
functions of $(\sqrt{m^2+P_t^2}\,)$~(b) and $(P_0+P_t)$~(c).}
  \label{fig:fig1}
\end{center}
\end{figure}

 To compare the quality of the data description by the 
forms~(\ref{Gazdzicki})-(\ref{Non-extensive})
we fitted these forms to the charged particle invariant cross
section measured in the UA1 experiment~\cite{UA1}. 
These data are selected as they cover a broad interval of~$P_t$ values
starting from very low~$P_t$.
The UA1 data are shown in Fig.~1(a) as function of~$P_t$.
Fig~1b and Fig~1c show the
relative differences between the data and the fit results
using form~(\ref{Gazdzicki}) with free mass parameter $m$ and
the form~(\ref{Power-law}) correspondingly.
 It is seen from the Fig.~1 that the form~(\ref{Gazdzicki}) doesn't
describe well the shape of the charged particle $P_t$ distribution, 
while the power-law form~(\ref{Power-law})
is in a good agreement with the data over almost whole~$P_t$ 
interval except of the very high~$P_t$.
Similarly good agreement with the data was found for the 
expression~(\ref{Non-extensive}), however, for simplicity,
we use the form~(\ref{Power-law}) only in the rest of the paper.
\begin{figure}[h]
\begin{center}
\hspace*{0.0cm}
\epsfig{
file=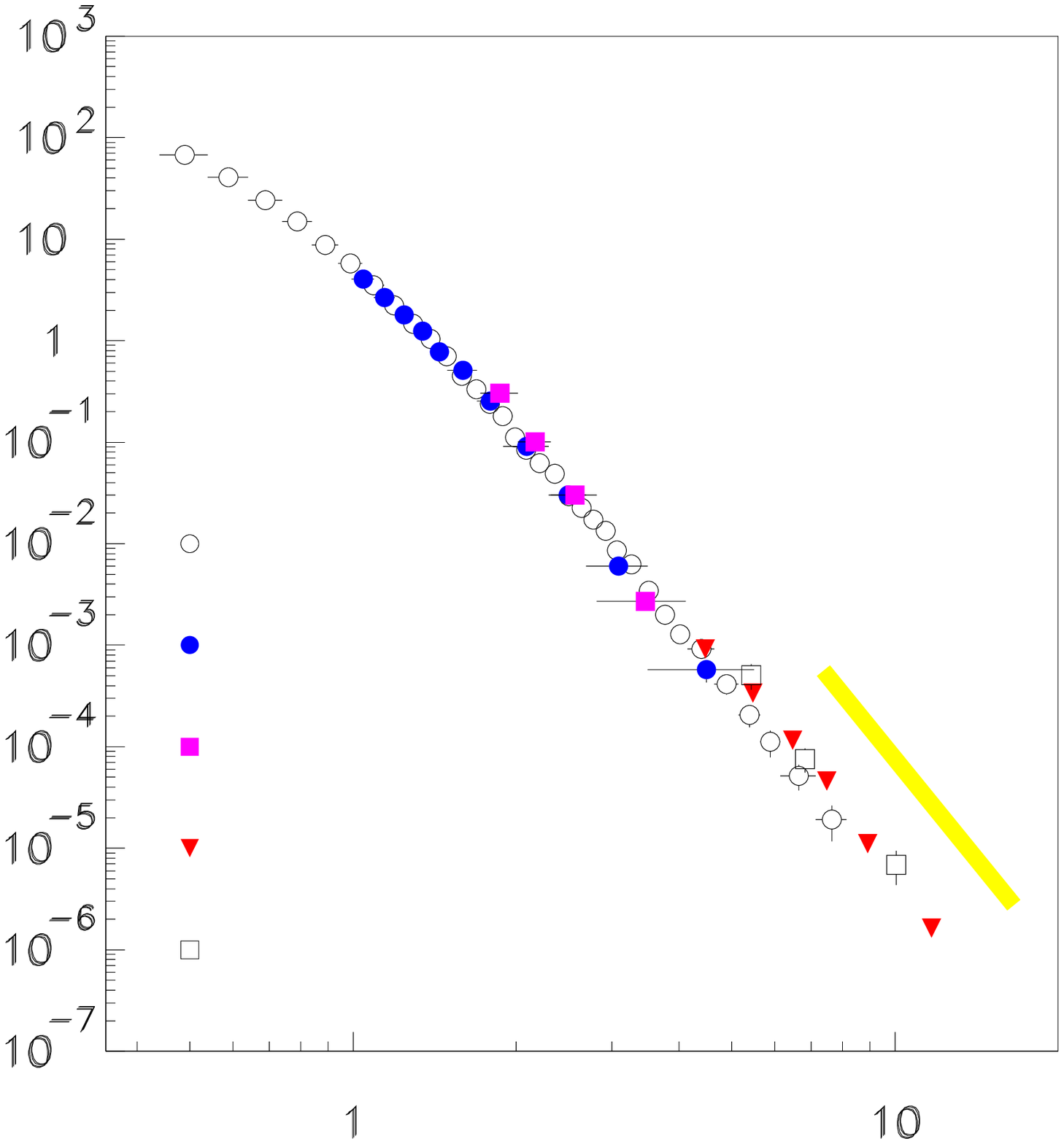,
        height=10.0cm,angle=0}
\put(-215,140){\large\bf $\pi$}
\put(-215,117){\large\bf $K$}
\put(-215,94){\large\bf $\Lambda$}
\put(-215,70){\large\bf $D^*$}
\put(-215,46){\large\bf $D_s$}
\put(-110,220){\Large\bf HERA}
\put(-145,200){\large\bf $\sqrt{s_{\gamma{p}}}=200~GeV$}
\put(-170,-5){\large\bf $m+P_t~[GeV]$}
\put(-305,50){\begin{sideways}\Large\bf
$\frac{1}{(2J+1)}E\left.\frac{d^3\sigma}{d^3p}\right|_{y_{lab}= 0}
[\mu{b}/GeV^2]$\end{sideways}}
\caption{
The invariant
cross-sections for different long-lived hadrons as
function of $m+p_T$, for rapidity
$y=0$ as measured in the laboratory system at H1 and ZEUS detectors.
The cross-sections are given for one spin and isospin projections.
The shaded bar corresponds to the $B$-meson cross section 
estimated from the measurement of inclusive $b\bar{b}$-production.
  \label{fig:incl2}}
\end{center}
\end{figure}

The power-law behavior of the data is a 
formal signature of the SOC pattern. 
The SOC model itself doesn't predict the
values of the parameters $P_0$ and $D$. These parameters are defined by
the underlying QCD dynamics.
The exponent $D$ has to match the pQCD calculations at large $P_t$.
Since QCD is flavor blind the exponent $D$ is likely to be the same
for different hadrons produced at the same collision conditions. 
If the SOC approach is applied to particle production,
$P_0$ defines a minimal scale at which the
$P_t$ scaling starts to be broken and is likely to be related
to the mass (as the only one dimensional macroscopic parameter available)
of the produced hadron. To test this hypothesis we compare
the spectra for different long-lived hadrons
assuming, for simplicity, $P_0$ in~(\ref{Power-law}) to be equal
to the hadron mass~$m$. 
A compilation of the invariant cross sections for   
$\pi^+$~\cite{zeus:pi}, $K^+$, $\Lambda$~\cite{h1:K},
$D_s^+$~\cite{zeus:Ds} and $D*^+$~\cite{zeus:D*} measured 
in photoproduction at HERA~$(\sqrt{s_{\gamma{p}}}\approx 200~GeV)$
is shown in Fig.2 as function of~$(m+P_t)$. 
To make the comparison of different species of hadrons
the cross sections are given for the particle's isospin and spin
projections. The pions were not identified in the HERA experiments
and have been recalculated from the measured charged particle spectrum
by reducing this spectrum by $40\%$ to take into account an admixture of
kaons, protons and long lived charged leptons.
Surprisingly, the spectra shown in Fig.2 for different hadrons can be 
described with a good approximation
by single power-law function\footnote{Recently, the H1 has 
reported~\cite{Ozerov} the preliminary proton photoproduction cross
section which supports the observed scaling.}. 
The production mechanism for different
long lived particles and for different intervals of $P_t$ is
self-similar, i.e. once established for charmed mesons, this mechanism is
valid for light pions also at low-$P_t$. 
We test further this universality with the high energy $p\bar{p}$
interaction.
In Fig.3 the invariant cross sections for $\pi^+$~\cite{cdf:pi}, 
$K^0$~\cite{cdf:K0}, $D^{*+}$~\cite{cdf:D*} and $B^+$~\cite{cdf:B} 
measured in $p\bar{p}$ collisions at the Tevatron~$(\sqrt{s}=1800~GeV)$
are presented as function of the $(m+P_t)$ scaling variable.
The Tevatron data show an approximate
scaling behavior for the $\pi$, $K$ and $D^*$ mesons
similar to that found at HERA $\gamma{p}$ collisions.
\begin{figure}[h]
\begin{center}
\hspace*{0.0cm}
\epsfig{
file=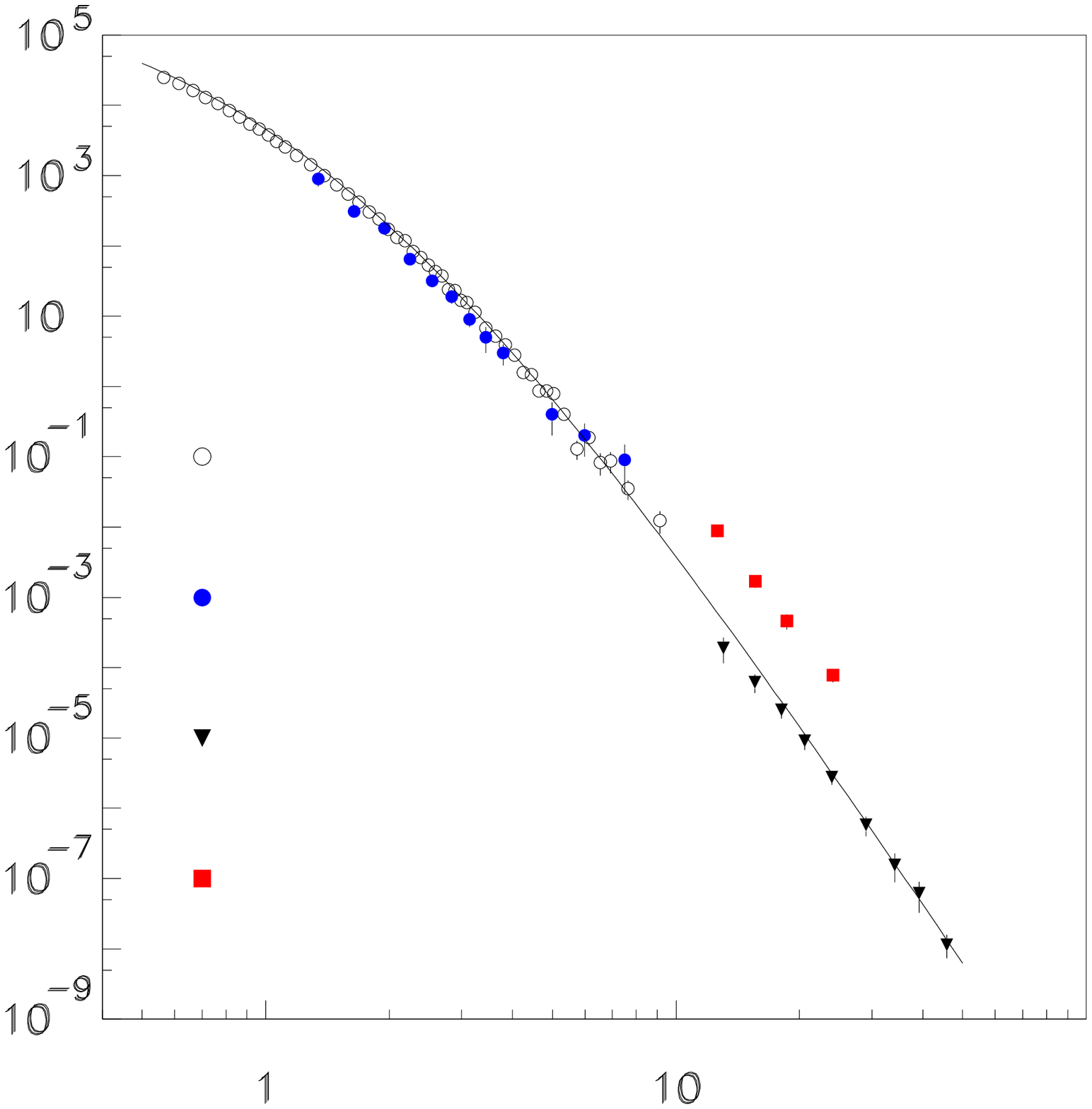,
        height=10.0cm,angle=0}
\put(-220,155){\large\bf $\pi$}
\put(-220,122){\large\bf $K$}
\put(-220,88){\large\bf $D^*$}
\put(-220,57){\large\bf $B^+$}
\put(-120,220){\Large Tevatron}
\put(-145,200){\large\bf $\sqrt{s_{p\bar{p}}}=1800~GeV$}
\put(-105,-5){\large\bf $m+P_t~[GeV]$}
\put(-315,50){\begin{sideways}\Large\bf
$\frac{1}{(2J+1)}E\left.\frac{d^3\sigma}{d^3p}\right|_{y_{lab}=0}
[\mu{b}/GeV^2]$\end{sideways}}
\caption{
The invariant
cross-sections for different long-lived hadrons as a
function of $m+p_T$, for rapidity
$y=0$ as measured in the laboratory system at CDF detector.
  \label{fig:incl3}}
\end{center}
\end{figure}

As seen from the Fig.~3 the
$B$-meson invariant cross section is a factor of $10$ higher 
than that of the lower mass hadrons when plotted at the same values of
$(m+P_t)$ variable. 
At HERA, though the $B$ mesons are not yet directly reconstructed,
the $B$-meson inclusive cross section was estimated using
the Lund model~\cite{Lund} with a normalization to the $b$-quark
production cross section measured at HERA~\cite{b-quark}. 
The result of these calculations is shown in Fig.2 with a shaded band. 
The band width is defined by the experimental uncertainty of the 
$b$-quark production cross section measurement. 
The calculated $B$-meson yield at HERA is again a
factor of $10$ higher than that expected for other hadrons shown in Fig.~2. 
Noteworthy, the QCD calculations 
for open beauty production strongly underestimate
the measured cross section at the Tevatron and HERA~\cite{bmore} 
as well. Recently, a new parton $k_t$-factorization approach was used
to describe the open beauty production~\cite{kt}, but it is still to be
demonstrated that this approach can be concistently applied to other
hadron spieces.
Within the statistical approach the observed excess of $B$-meson
production over the universal behaviour of the scale dependence of other
mesons signals a different dynamics or the existence of an additional
mechanism of $B$-production.

In summary, we noted that particle production in high energy 
hadronic collisions has the formal properties of a Self-Organized
Critical process, name- ly:
a) the hadronizing partons are known to represent an open
strongly interacting dynamical system far from equilibrium -- a
typical system to deal with the SOC;
b) the resulting hadronic state shows long-range particle-particle
correlations;
c) the particle spectrum is described by a power-law distribution
as function of transverse momentum and mass of the particle;
d) this distribution is the same for different long-lived hadrons
and is defined by macroscopic parameters like mass and spin of the
particle as expected for statistical models.
Whether the latter is a pure coincidence or a demonstration of a
statistical (alike SOC) nature of the hadron production is an open
question. To prove an eligibility of the SOC
mechanism in particle production, a successful quantum cellular automaton
describing a quark-gluon system has to be built. It is possibly could be
done in the future using QCD lattice calculations. 

For completeness we give also the earlier applications of the SOC in
high energy physics known to us. The SOC was suggested~\cite{Boros} to
describe inelastic diffractive scattering. Fractal properties of
the hadronic collisions have been discussed in~\cite{Tokarev}.
In~\cite{Lastovicka} a concept of fractal dimensions of the proton
structure function was introduced. This gave an excellent description of
the low-$x$ HERA data, both in the non-perturbative and the
deep-inelastic domain. These observations together with the
self-similarity in particle production discussed in this paper make the
SOC approach to be a candidate for an effective
picture to describe a variety of non-perturbative QCD phenomena.

\section*{Acknowledgments}
The work was partially supported by
Russian Foundation for Basic Research, grant 
RFBR-01-02-16431 and grant RFBR-00-15-96584.

\end{document}